\documentclass[twocolumn,floatfix,preprintnumbers,amsmath,amssymb,prb]{revtex4}
\usepackage{graphicx}
\usepackage{dcolumn}
\usepackage{latexsym}
\usepackage{epsfig,color}
\usepackage{epstopdf}

\newcommand{\be}{\begin{equation}}
\newcommand{\ee}{\end{equation}}
\newcommand{\bea}{\begin{eqnarray}}
\newcommand{\eea}{\end{eqnarray}}

\newcommand{\p}{\partial}
\newcommand{\s}{\sigma}

\newcommand{\la}{\langle}
\newcommand{\ra}{\rangle}
\newcommand{\rd}{\mbox{d}}
\newcommand{\ri}{\mbox{i}}

\newcommand{\eps}{\epsilon}
\newcommand{\nn}{\nonumber}

\newcommand{\bz}{\bar{z}}
\newcommand{\bw}{\bar{w}}

\newcommand{\vare}{\varepsilon}

\newcommand{\vxi}{\mbox{\boldmath $\xi$}}
\newcommand{\veta}{\mbox{\boldmath $\eta$}}
\newcommand{\vtau}{\mbox{\boldmath $\tau$}}

\newcommand{\vrho}{\mbox{\boldmath $\rho$}}

\newcommand{\vDelta}{\mbox{\boldmath $\Delta$}}

\newcommand{\vll}{\mbox{\boldmath $\ll$}}

\begin{document}
\input{epsf}
\title{Quantum phase transitions in a strongly entangled spin-orbital chain:
\\A field-theoretical approach}
\author{Alexander Nersesyan}
\affiliation{
The Abdus Salam International Centre for Theoretical Physics, 34100, Trieste, Italy \\
Andronikashvili Institute of Physics, Tamarashvili 6, 0177, Tbilisi, Georgia\\
Center of Condensed Mater Physics, ITP, Ilia State University, 0162, Tbilisi, Georgia}
\author{Gia-Wei Chern}
\author{Natalia  B. Perkins}
\affiliation{Department of Physics, University of Wisconsin, Madison, Wisconsin 53706, USA}

\begin{abstract}
Motivated by recent experiments on quasi-1D vanadium oxides, we
study quantum phase transitions in a one-dimensional spin-orbital
model describing a Haldane chain and a classical Ising chain locally
coupled by the relativistic spin-orbit interaction. By employing a
field-theoretical approach, we analyze the topology of the
ground-state phase diagram and identify the nature of the phase
transitions. In the strong coupling limit, a long-range N\'eel order
of entangled spin and orbital angular momentum appears in the ground
state. We~find that, depending on the relative scales of the spin
and orbital gaps, the linear chain follows two distinct routes to
reach the N\'eel state. First, when the orbital exchange is the
dominating energy scale, a two-stage ordering takes place in which
the magnetic transition is followed by melting of the orbital Ising
order; both transitions belong to the two-dimensional Ising
universality class. In the opposite limit, the low-energy orbital
modes undergo a continuous reordering transition which represents a
line of Gaussian critical points. On this line the orbital degrees
of freedom form a Tomonaga-Luttinger liquid. We argue that the
emergence of the Gaussian criticality results from merging of the
two Ising transitions in the strong hybridization region where the
characteristic spin and orbital energy scales become comparable.
Finally, we show that, due to the spin-orbit coupling, an external
magnetic field acting on the spins can induce an orbital Ising
transition.
\end{abstract}
\maketitle

\section{Introduction}

Over the past decades, one-dimensional spin-orbital models have been
a subject of intensive theoretical studies. The interest is to a
large extent motivated by experimental discovery of unusual magnetic
properties in various quasi-one-dimensional Mott insulators.
\cite{axtell,isobe} The inter-dependence of spin and orbital degrees
of freedom is usually described by the so-called Kugel-Khomskii
Hamiltonian in which the effective spin exchange constant depends on
the orbital configuration and vice versa. \cite{kk, tokura} Another
mechanism of coupling spin and orbital degrees of freedom is the
on-site relativistic spin-orbit (SO) interaction $\lambda\, \mathbf
L\cdot\mathbf S$, where $\mathbf L$ is the orbital angular momentum
and $\lambda$ is the coupling constant. In compounds with quenched
orbital degrees of freedom, the presence of the SO term usually
leads to the single-ion spin anisotropy $D S_z^{\,2}$ where $D \sim
\lambda^2/\Delta$ and $\Delta$ denotes the energy scale of the
crystal field which lifts the degenerate orbital states.

For systems with residual orbital degeneracy, on the other hand, the
effect of the SO term is much less explored compared with the
Kugel-Khomskii-type coupling. Due to the directional dependence of
the orbital wave functions, the SU(2) symmetry of the Heisenberg
spin exchange is expected to be broken in the presence of the SO
interaction. The resultant spin anisotropy is likely to induce a
long-range magnetic order in the spin sector. A more intriguing
question is what happens to the orbital sector. To answer this
question, one needs to consider the details of the interplay between
the orbital exchange and the SO coupling. Here we consider the
simplest case of
a two-fold orbital degeneracy per site. Specifically,
the two degenerate states could be the $d_{yz}$ and $d_{zx}$
orbitals in a tetragonal crystal field observed in several
transition-metal compounds. We introduce pseudospin-1/2 operators
$\tau^a$ ($a=x,y,z$) to describe the doublet orbital degrees of
freedom assuming that $\tau^z = \pm 1$ correspond to the states
$|yz\rangle$ and $i|zx\rangle$, respectively. Alternatively, one can
also realize the double orbital degeneracy in the Mott-insulating
phase of a 1D fermionic optical lattice where the eigenvectors of
$\tau^z$ refers to $p_x$ and $p_y$ orbitals in an anisotropic
potential.\cite{zhao,wu} Restricted to this doublet space, the
orbital angular momentum operator
$\mathbf L = (0, 0,
\tau^x)$. This can be easily seen by noting that the eigenstates of
$\tau^x$ carry an angular momentum $\langle L^z \rangle = \pm 1$.

The exchange interaction between localized orbital degrees of
freedom is characterized by its highly directional dependence: the
interaction energy only depends on whether the relevant orbital is
occupied for bonds of a given orientation. This is particularly true
for interactions dominated by direct exchange mechanism. Denoting
the relevant orbital projectors on a given bond as $P = (1 +
\tau^\beta)/2$, where $\tau^\beta /2$ is an appropriate pseudospin-1/2
operator ($\tau^{\beta}$ being a Pauli matrix), the orbital interaction is thus described by an Ising-type
term $\tau^{\beta}_i\tau^{\beta}_j$. The well studied orbital
compass model and Kitaev model both belong to this
category.\cite{nussinov,kitaev} The quantum nature of these models
comes from the fact that different operators $\tau^\beta$, which do
not commute with each other, are used for bonds of different types.
To avoid unnecessary complications coming from the details of
orbital interactions, we assume that there is only one type of bond
in our 1D system and the orbital interaction is thus governed by a
classical Ising Hamiltonian.

We incorporate these features into the following toy model of
spin-orbital chain ($J_s, J_{\tau} > 0$):
\begin{eqnarray}
    \label{ham}
    H &=& H_S + H_{\tau} + H_{S\tau} \\
    &=&
    J_S\sum_n \mathbf S_n\cdot\mathbf S_{n+1} + J_{\tau}\sum_n
    \tau^z_n\tau^z_{n+1} + \lambda \sum_n \tau^x_n S^z_n. \nonumber
\end{eqnarray}
Motivated by the recent experimental characterizations of quasi-1D
vanadium oxides,\cite{mamiya,reehuis,onoda,lee,pieper,niazi}
here we focus on the case of quantum spin with length $S = 1$. The
above model thus describes a Haldane chain locally coupled to a
classical Ising chain by the SO interaction $H_{S\tau}$. The role of
the $\lambda$-term is two-fold: firstly it introduces anisotropy to
the spin-1 subsystem, and secondly it endows quantum dynamics to the
otherwise classical Ising chain.

Before turning to a detailed study of the phase diagram of
model~(\ref{ham}), we first discuss its connections to real
compounds. As mentioned above, the interest in the toy model is
partly motivated by the recent experimental progress on vanadium
oxides which include spinel ZnV$_2$O$_4$
\cite{mamiya,reehuis,onoda,lee} and quasi-1D CaV$_2$O$_4$.
\cite{pieper,niazi} In both types of vanadates, the two $d$
electrons of V$^{3+}$ ions have a spin $S = 1$ in accordance with
Hund's rule. In the low-temperature phase of both vanadates, the
vanadium site embedded in a flattened VO$_6$ octahedron has a
tetragonal symmetry. This tetragonal crystal field splits the
degenerate $t_{2g}$ triplet into a singlet and a doublet. As one of
the two $d$ electrons occupies the lower-energy $d_{xy}$ state, a
double orbital degeneracy arises as the second electron could occupy
either $d_{zx}$ or $d_{yz}$ orbitals. The fact that the $d_{xy}$
orbital is occupied everywhere also contributes to the formation of
weakly coupled quasi-1D spin-1 chains in these compounds.
\cite{chern1} On the other hand, the details of the orbital exchange
depends on the geometry of the lattice and in the case of vanadium
spinel the orbital interaction is of three-dimensional nature. The
Ising orbital Hamiltonian in Eq.~(\ref{ham}) thus should be regarded
as an effective interaction in the mean-field sense. Nonetheless,
the toy model provides a first step towards understanding the
essential physics introduced by the SO coupling. Moreover, many
conclusions of this paper can be applied to the case of quasi-1D
compound CaV$_2$O$_4$ where the vanadium ions form a zigzag chain.

It is instructive to first establish regions of stable massive
phases. In the decoupling limit, $\lambda \to 0$, our model
describes two gapped systems: a quantum spin-1 Heisenberg chain and
a classical orbital Ising chain. The ground state of the spin sector
is a disordered quantum spin liquid with a finite spectral gap\cite{haldane}
$\Delta_S$, whereas the orbital ground state is characterized by a
classical N\'{e}el order along the chain: $\langle \tau^z_n \rangle
= (-1)^n\,\eta^z$. Quantum effects in the orbital sector induced by
the SO coupling play a minor role. Obviously, just because of being
gapped, both the spin-liquid phase and the orbital ordered state are
stable as long as $\lambda$ remains small. Consider now the opposite
limit, $\lambda \gg J_{S}, J_{\tau}$. In the zeroth order
approximation, the model is dominated by the single-ion term
$H_{S\tau}$ whose doubly degenerate eigenstates $|\pm \rangle = |S^z
= \pm 1\rangle \otimes |\tau^x = \pm 1\rangle$ represent locally
entangled spin and orbital degrees of freedom. Switching on small
$J_S$ and $J_{\tau}$ leads to a staggered ordering of the
$|+\rangle$ and $|-\rangle$ states along the chain. Physically, the
large-$\lambda$ ground state can be viewed as a simultaneous N\'eel
ordering of spin and orbital angular momentum characterized by order
parameters $\zeta$ and $\eta^x$ such that $\langle S^z_n \rangle =
(-1)^n \zeta$ and $\langle L^z_n \rangle = \langle \tau^x_n \rangle
= (-1)^n \eta^x$. The Ising order parameter $\eta^z$ vanishes
identically in this phase.

These observations naturally lead to the following questions. How is
the magnetically ordered N\'eel state at large $\lambda$ connected
to the disordered Haldane phase as $\lambda \to 0$ ? What is the
scenario for the orbital reorientation transition $\eta^z \to
\eta^x$, which is of essentially quantum nature ? In this paper we
employ the field-theoretical approach to address these questions. We
first note that the one-dimensional model (\ref{ham}) is not exactly
integrable. As a consequence, the regime of strong hybridization of
the spin and orbital excitations, which is the case when $J_{\tau}$,
$J_S$ and $\lambda$ are all of the same order, stays beyond the
reach of approximate analytical methods. We thus will be mainly
dealing with limiting cases $J_{\tau} \gg J_S$ and $J_{\tau} \ll
J_S$, in which one can integrate out the ``fast'' variables to
obtain an effective action for the ``slow'' modes. Following this
approach, we establish the topology and main features of the
ground-state phase diagram in the accessible parts of the parameter
space of the model. We were able to unambiguously identify the
universality classes of quantum criticalities separating different
massive phases. Using plausible arguments we comment on some
features of the model in the regime of strong spin-orbital
hybridization.

We demonstrate that the aforementioned reorientation transition
$\eta^z \to \eta^x$ can be realized in one of two possible ways. In
the limit of large $J_{\tau}$, we find a sequence of two quantum
Ising transitions and an intermediate massive phase, sandwiched
between these critical lines, in which both $\eta^z$ and $\eta^x$
are nonzero. This is consistent with the recent findings
\cite{chern2} based on DMRG calculations and some analytical
estimations. In the opposite limit, when the Haldane gap $\Delta_S$
is the largest energy scale, integrating out the spin excitations
yields an effective lowest-energy action for the orbital degrees of
freedom, which shows that the $\eta^z \to \eta^x$ crossover takes
place as a single Gaussian quantum criticality. At this critical
point, the orbital degrees of freedom display an extremely quantum
behaviour: they are gapless and form a Tomonaga-Luttinger liquid.
This is the main result of this paper. We bring about arguments
suggesting that the emergence of the Gaussian critical line is the
result of merging of the two Ising criticalities in the region of
strong spin-orbital hybridization.

\medskip

Any field-theoretical treatment of the model (\ref{ham}) must be
based on a properly chosen contiuum description of the spin-1
antiferromagnetic Heisenberg chain. Its properties have been
thoroughly studied, both analytically and numerically (see for a
recent review Ref.~\onlinecite{mk}). In what follows, the spin
sector of the model (\ref{ham}) will be treated within the
O(3)-symmetric Majorana field theory, proposed by Tsvelik: \cite{t}
\bea {\cal H}_{\rm M} = \sum_{a=1,2,3}\left[ \frac{\ri v}{2} \left(
\xi^a _L \p_x \xi^a _L - \xi^a _R \p_x \xi^a _R \right) - \ri
m \xi^a _R \xi^a _L \right] + {\cal H}_{\rm int}. \nonumber \\
\label{Maj-Tsv} \eea Here $\xi^a _{R,L} (x)$ is a degenerate triplet
of real (Majorana) Fermi fields with a mass $m$, the indices $R$ and
$L$ label the chirality of the particles, and
\[
{\cal H}_{\rm int} = \frac{1}{2} g \sum_{a} \left( \xi^a _R \xi^a _L
\right)^2
\]
is a weak four-fermion interaction which can be treated
perturbatively. The continuum theory (\ref{Maj-Tsv}) adequately
describes the low-energy properties of the generalized spin-1
bilinear-biquadratic chain \be H_S \to \bar{H}_S = J_S \sum_n \left[
{\bf S}_n \cdot {\bf S}_{n+1} - \beta \left( {\bf S}_n \cdot {\bf
S}_{n+1} \right)^2  \right].\label{biquad-chain} \ee in the vicinity
of the critical point $\beta = 1$. \cite{babu} This quantum
criticality belongs to the universality class of the SU(2)$_2$
Wess-Zumino-Novikov-Witten (WZNW) model with central charge $c=3/2$.
\medskip

At small deviations from criticality the Majorana mass $m \sim  J_S
|\beta - 1|$ determines the magnitude of the triplet gap, $\Delta_S
= |m| \ll J_S$. The theory of a massive triplet of Majorana fermions
is equivalent to a system of three degenerate noncritical 2D Ising
models, with $m \sim (T-T_c)/T_c$. This is one of the most appealing
features of the theory because the most strongly fluctuating
physical fields of the $S=1$ chain, namely the staggered
magnetization and dimerization operators, have a simple local
representation in terms of the Ising order and disorder
parameters.\cite{t,snt,gnt} It is this fact that greatly simplifies
the analysis of the spin-orbital model (\ref{ham}). While the
correspondence between the models (\ref{Maj-Tsv}) and
(\ref{biquad-chain}) is well justified at $|\beta - 1| \ll 1$, it is
believed that the Majorana model (\ref{Maj-Tsv}) captures generic
properties of the Haldane spin-liquid phase of the spin-1 chain,
even though at large deviations from criticality ($|\beta - 1| \sim
1, ~\Delta_S \sim J_s$) all  parameters of the model should be
treated as phenomenological.

\medskip

The remainder of the paper is organized as follows. We start our
discussion with Sec.~\ref{maj_intro} which contains a brief summary
of known facts about the Majorana model\cite{t} that will be used in
the rest of the paper. In Sec.~\ref{heavy_orbit} we consider the
limit $J_{\tau}/\Delta_S \gg 1$ and by integrating out the `fast'
orbital modes, show that on increasing the SO coupling $\lambda$ the
system undergoes a sequence of two consecutive quantum Ising
transitions in the spin and orbital sectors, respectively. In
section \ref{gauss} we analyze the opposite limiting case,
$J_{\tau}/\Delta_S \ll 1$, and, by integrating over the `fast' spin
modes, show that there exists a single Gaussian transition in the
orbital sector accompanied by a Neel ordering of the spins. We then
conjecture on the topology of the ground-state phase diagram of the
model.  In Sec.~\ref{V} we  show that spin-orbital hybridization effects
near the orbital Gaussian transition  lead to the appearance of a non-zero
 spectral weight well below the Haldane gap which can be detected by
inelastic neutron scattering experiments and NMR measurements.
In Sec.~\ref{mag_field} we comment on the role of an
external magnetic field. We show that, through the SO interaction, a
sufficiently strong magnetic field affects the orbital degrees of
freedom and can lead to a quantum Ising transition in the orbital
sector. Sec.~\ref{con} contains a summary of the obtained results
and conclusions. The paper has two appendices containing certain
technical details.

\medskip

\begin{figure}
\begin{center}
\epsfig{file=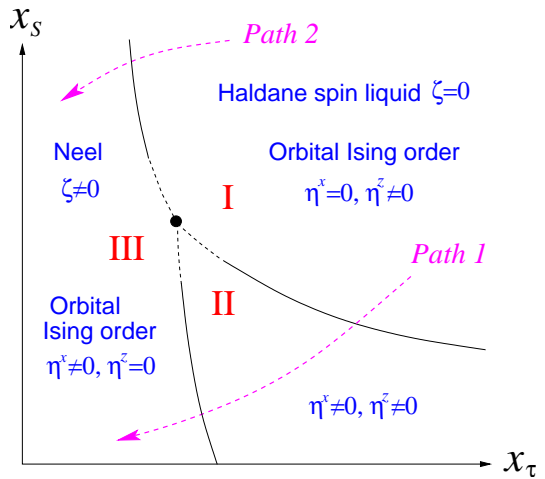,width=7cm}
\end{center}
\caption{Schematic phase diagram of the model on the ($x_S$,
$x_{\tau}$)-plane, where $x_S = \Delta_S/\lambda$ and $x_{\tau} =
J_{\tau}/\lambda$.} \label{merging}
\end{figure}


\section{Some facts about Majorana theory of spin-1 chain}\label{maj_theory}
\label{maj_intro}

In this Section, we provide some details about the O(3)-symmetric
Majorana field theory,\cite{t} Eq.~(\ref{Maj-Tsv}), which represents
the continuum limit of the biquadratic spin-1 model
(\ref{biquad-chain}) at $|\beta - 1| \ll 1$.
\medskip

In the continuum description, the local spin density of the spin
model (\ref{biquad-chain}) has contributions from the low-energy
modes centered in momentum space at $q = 0$ and $q=\pi$: \be {\bf
S}(x) = {\bf I}_{R} (x) + {\bf I}_{L} (x) + (-1)^{x/a_0} {\bf N}(x)
\label{spin-den-ladder} \ee
The smooth part of the local
magnetization, ${\bf I} = {\bf I}_R + {\bf I}_L$, is a sum of the
level-2 chiral vector currents. The SU(2)$_2$ Kac-Moody algebra of these currents is faithfully
reproduced in terms of a triplet of massless Majorana fields \cite{zamo_fat}
$\vxi = (\xi^1,\xi^2,\xi^3)$:
\be {\bf I}_{\nu} =
- \frac{i}{2} \left( \vxi_{\nu} \times \vxi_{\nu} \right),
~~~(\nu=R,L) \label{vector_currents} \ee
This fact is not surprising because, as already mentioned, the central charge
of the SU(2)$_2$ WZNW theory is $c=3/2$, whereas that of the theory of a massless
Majorana fermion (equivalently, critical 2D Ising model) is $c=1/2$.
At small deviations from criticality ($|\beta - 1| \ll 1$)
the fermions acquire a mass.
Strongly fluctuating fields
of the spin-1 chain, the staggered magnetization ${\bf N} (x)$ and
dimerization operator $\eps (x) = (-1)^n {\bf S}_{n} \cdot {\bf
S}_{n+1}$, are nonlocal in terms of the Majorana fields but admit a
simple representation in terms of the order, $\s$, and disorder,
$\mu$, operators of the related noncritical Ising models:
\bea {\bf N} &\sim & (1/\alpha)
\left( \s_1 \mu_2 \mu_3 ,  ~\mu_1 \s_2 \mu_3 , ~\mu_1 \mu_2 \s_3
\right),\nonumber \\ \eps &\sim & (1/\alpha)\, \s_1 \s_2 \s_3 ,
\label{N-eps} \eea where $\alpha \sim a_0$ is a short-distance
cutoff of the continuum theory. These expressions together with
their duals (i.e. their counterparts obtained by the duality transformation
in all Ising copies, $\s_a \leftrightarrow \mu_a $) determine the vector
and scalar parts of the WZNW 2$\times$2 matrix field $\hat{g}$
which is a primary scalar field with scaling dimension 3/8.
It has been demonstrated in
Ref. \onlinecite{zamo_fat} that using the representation (\ref{N-eps})
and the short-distance operator product expansions for the Ising fields,
one correctly reproduces all fusion rules of the SU(2)$_2$ WZNW model.
An equivalent way to make sure that this is indeed the case is to
consider the four-Majorana representation of the weakly coupled
spin-1/2 Heisenberg ladder \cite{snt,gnt} and take the limit
of a infinite singlet Majorana mass to map the low-energy sector
of the model on the O(3) theory (\ref{Maj-Tsv}).
\medskip

In the spin-liquid phase of the spin
chain (\ref{biquad-chain}), which is the case $\beta < 1$, the
Majorana mass $m$ is positive, implying that the degenerate triplet
of 2D Ising models is in a disordered phase: $\la \s_a \ra = 0$,
$\la \mu_a \ra \neq 0$ $~(a=1,2,3)$. In particular, this implies
that the O(3) symmetry remains unbroken, $\la  {\bf N}\ra = 0$, and
the ground state of the system is not spontaneously dimerized, $\la
\eps \ra = 0$.
\medskip

The representation (\ref{N-eps}) proves to be very useful for
calculating the dynamical spin correlation functions because the
asymptotics of the Ising correlators $\la \s(x,\tau) \s (0,0) \ra$
and $\la \mu(x,\tau) \mu (0,0) \ra$ are well known both at
criticality and in a noncritical regime. In the disordered phase ($m
> 0$), the leading asymptotics of the Ising correlators are: \bea \la
\mu ({\bf r}) \mu ({\bf 0})\ra &\sim & \left( a/\xi_S \right)^{1/4}
\left[ 1 + O(e^{-2r/\xi_S}) \right], \nonumber \\ \la \s ({\bf r})
\s ({\bf 0})\ra &\sim &\left( a/\xi_S \right)^{1/4} \sqrt{\xi_S/r}
~e^{-r/\xi_S} \label{ising-corr} \eea where $\xi_S = v/m$ is the
correlation length, and $r = \sqrt{x^2 + v^2 \tau^2}$. (By duality,
in the ordered phase ($m<0$) the asymptotics of the correlators in
(\ref{ising-corr}) must be interchanged.) Correspondingly, the
dynamical correlation function \be \la {\bf N}({\bf r}) {\bf N}
({\bf 0}) \ra \sim \left( a/\xi_S \right)^{3/4} \sqrt{\xi_S /r}
~e^{-r/\xi_S}. \label{NN} \ee Its Fourier transform at $q \sim \pi$
and small $\omega$ describes a coherent excitation -- a triplet
magnon with the mass gap $m$: \be \Im m~\chi(q, \omega) \sim
\frac{m}{|\omega|} \delta \left( \omega - \sqrt{(q - \pi)^2 v^2 +
m^2}\right). \label{magnon_prop} \ee

Since the single-ion anisotropy
$
H_{\rm anis} = D \sum_n \left( S^z _n \right)^2
$
lowers the original O(3) symmetry down to O(2) $\times$ Z$_2$,
one expects \cite{t} that in the continuum theory
it will induce anisotropy in the Majorana masses
\[
m_1 = m_2 \neq m_3,
\]
as well as in the coupling constants parametrizing the four-fermion interaction:
\[
{\cal H}_{\rm int} \to \frac{1}{2} \sum_{a\neq b} g_{ab} \left( \xi^a _R \xi^a _L \right)
\left( \xi^b _R \xi^b _L \right), ~~~g_{13} =  g_{23}\neq g_{12}~.
\]
 This can be checked
by using the correspondence (\ref{spin-den-ladder}) and
short-distance operator product expansions (OPE) for the physical
fields. There will also appear anisotropy in the velocities, $v_1 =
v_2 \neq v_3$,   but we will systematically neglect this effect.
Thus, we have $H_{\rm anis} = \int \rd x~ {\cal H}_{\rm anis}$, with
\bea {\cal H}_{\rm anis} = D \alpha \int \rd x \left[ I^3 (x) I^3
(x+\alpha) +   N^3 (x)  N^3 (x+\alpha) \right], \label{anis_cont}
\quad \eea where $\alpha \sim a$ is a short-distance cutoff of the
continuum theory. Using (\ref{vector_currents}) and keeping only the
Lorentz invariant terms (i.e. neglecting renormalization of the
velocities) we can replace $(I^3)^2$ by $2 I^3 _R I^3 _L$. To treat
the second term in the r.h.s. of (\ref{anis_cont}), we need OPEs for
the products of Ising operators: \cite{cft} \bea & & \s(z, \bz) \s
(w, \bw) \nn \\  & & \,\, = \frac{1}{\sqrt{2}}\left(
\frac{\alpha}{|z-w|} \right)^{1/4} \bigl[1 - \pi |z-w| \vare (w,
\bw) \bigr], \label{ope1} \eea \bea & & \mu(z, \bz) \mu (w, \bw) \nn \\
& & \,\,= \frac{1}{\sqrt{2}}\left( \frac{\alpha}{|z-w|}
\right)^{1/4} \bigl[1 + \pi |z-w| \vare (w, \bw) \bigr].
\label{ope2} \eea Here $\vare = i \xi_R \xi_L$ is the energy density
(mass bilinear) of the Ising model,
 $z = v \tau + \ri x$ and $w = v \tau' + \ri x' $
are two-dimensional complex coordinates, $\bz$ and $\bw$ are their conjugates.
 From the above OPEs it follows that
\bea & & N^3 (x)  N^3 (x+\alpha) = \ri (\pi /\alpha) \left(  \xi^1
_R \xi^1 _L +   \xi^2 _R \xi^2 _L -
\xi^3 _R \xi^3 _L \right)\nn \\
& & \quad - (\pi^2 C) [(\xi^1 _R \xi^1 _L)(\xi^2 _R \xi^2 _L) -
(\xi^1 _R \xi^1 _L)(\xi^3 _R \xi^3 _L) - (\xi^2 _R \xi^2 _L)(\xi^3
_R \xi^3 _L)], \nn \eea where $C \sim 1 $ is a nonuniversal
constant. As a result, \bea {\cal H}_{\rm anis} = - \ri
\sum_{a=1,2,3} \delta m_a~ \xi^a _R \xi^a _L + \frac{1}{2}
\sum_{a\neq b} \delta g_{ij} \left( \xi^a _R \xi^a _L \right)\left(
\xi^b _R \xi^b _L \right), \quad \label{anis_tot_cont} \eea where
\be \delta m_1 = \delta m_2 = - \delta m_3 = - (\pi C)D
\label{mass_corr} \ee are corrections to the single-fermion masses,
and $ \delta g_{12} = (2 - \pi^2 C )D \alpha, $ $ \delta g_{13} =
\delta g_{23} = \pi^2 C D \alpha $ are coupling constants of the
induced interaction between the fermions. Smallness of the Majorana
masses ($|m|\alpha/v \ll 1$) implies that the additional mass
renormalizations caused by the interaction in (\ref{anis_tot_cont})
are relatively small, $m (D\alpha/v) \ln (v/|m|\alpha) \ll D$, so
that the main effect of the single-ion anisotropy is the additive
renormalization of the fermionic masses, $m_a = m + \delta m_a$,
with $\delta m_a$ given by Eq.(\ref{mass_corr}).
\medskip

The cases $D > 0$ and $D<0$ correspond to an easy-plane and
easy-axis anisotropy, respectively. The spin anisotropy
(\ref{spin-anis}) induced by the spin-orbit coupling is of the
easy-axis type. At $D < 0$ the singlet Majorana fermion, $\xi^3$, is
the lightest, $m_3 < m_1 = m_2$. Increasing anisotropy drives the
system towards an Ising criticality at $D = - D^*$, where $m_3 = 0$.
At $D < - D^*$ the system occurs in a new phase where the Ising
doublet remains disordered while the singlet Ising system becomes
ordered. It then immediately follows from the representation
(\ref{N-eps}) that the new phase is characterized by a N\'eel
long-range order with $\la  N^3 \ra \neq 0$. Transverse spin
fluctuations, as well as fluctuations of dimerization, are
incoherent in this phase.

\section{Two Ising transitions in the $\vDelta_{\bf S} \vll {\bf J}_{\vtau}$ limit }
\label{heavy_orbit}

Now we turn to our model (\ref{ham}).
Let us consider the case when, in the absence of spin-orbit
coupling, the orbital  gap is the largest: $J_{\tau} \gg J_s$. The
orbital pseudospins then represent the `fast' subsystem and can be
integrated out. Assuming that $\lambda \ll J_{\tau}$, we treat the
spin-orbit coupling perturbatively.  In this case, the zero order
Hamiltonian $H_0 = H_S + H_{\tau}$ describes decoupled spin and orbital
systems,  while the spin-orbit interaction $H_{S\tau}$ denotes perturbation.
Defining the interaction representation for all operators according
to $A(\tau) = e^{\tau H_0} A e^{- \tau H_0}$ (here $\tau$ denotes
imaginary time), the interaction term in the Euclidian action is
given by
\begin{eqnarray}
    S_{S\tau} = \lambda \sum_n \int  {\rm d} \tau~ \tau^x _{n}(\tau) S^z_n (\tau).
\end{eqnarray}
The first nonvanishing correction to the effective action in the spin sector
is of the second order in $\lambda$:
\begin{eqnarray}
    \label{eff_action1}
    \Delta S_{s}  = \frac{-\lambda^2}{2}
    \sum_{nm} \int {\rm d}\tau_1 {\rm d}\tau_2
    \bigl\langle \tau^x_n(\tau_1) \tau^x_m(\tau_2) \bigr\rangle_{\!\tau}
    S_n^{z}(\tau_1) S^{z}_m (\tau_2).~\nn \!\!\!\!\!\!\\
\end{eqnarray}
Averaging in the right-hand side of (\ref{eff_action1}) goes over
configurations of the classical Ising chain $H_{\tau}$. The
correlation function $\langle \tau^x _n (\tau_1) \tau^x _m  (\tau_2)
\rangle_{\tau}$ is calculated in  Appendix \ref{ising_corr}. It
is spatially ultralocal (because there are no propagating
excitations in the classical Ising model) and rapidly decaying at
the characteristic time $\sim 1/J_{\tau}$, which is much shorter than
the spin correlation time $\sim 1/\Delta_0$:
\begin{eqnarray}
    \langle \tau^x _n (\tau_1) \tau^x _m  (\tau_2) \rangle_{\tau}
    = \delta_{nm} \exp \left( - 4 J_{\tau} |\tau_1 - \tau_2|
    \right).
\end{eqnarray}
Passing to new variables, $\tau = (\tau_1 + \tau_2)/2$ and $\rho =
\tau_1 - \tau_2$, and integrating over $\rho$ yields a correction to
the effective spin action which has the form of a single-ion spin
anisotropy. Thus in the second order in $\lambda$, the spin
Hamiltonian acquires an additional term
\begin{equation}
    H_{\rm ani} = -  \frac{\lambda^2}{4J_{\tau}}
    \sum_n \left( S^z _n \right)^2 . \label{spin-anis}
\end{equation}
The anisotropy splits the Majorana triplet into a doublet $(\xi^1,
\xi^2)$ and singlet $(\xi^3)$, with masses
\begin{eqnarray}
    m_1 =
    m_2 = m + \frac{\pi C\lambda^2 }{4 J_{\tau}} , ~~~ m_3 = m -
    \frac{\pi C\lambda^2 }{4 J_{\tau}},\label{masses_split}
\end{eqnarray}
where $C \sim 1$ is a nonuniversal positive constant. The anisotropy
is of the easy-axis type, so that the singlet mode has a smaller
mass gap.

As long as all the masses $m_a$ remain positive, the system
maintains the properties of an anisotropic Haldane's spin-liquid.
The dynamical spin susceptibilities calculated at small $\omega$ and
$q \sim \pi$ (see Sec. \ref{maj_intro}), \bea &&\Im m~ \chi^{xx}(q,
\omega) =  \Im m~\chi^{yy}(q, \omega) \\ &&
\quad\quad\quad\quad\quad\quad\sim
\frac{m_1}{|\omega|}\delta \left(\omega - \sqrt{(q - \pi)^2 v^2 + m^2 _1}\right), \nn\\
&&\Im m~\chi^{zz}(q, \omega) \sim \frac{m_3}{|\omega|}\delta\left(
\omega - \sqrt{ (q - \pi)^2 v^2 + m^2 _3}\right),\nn \label{chi-s}
\eea indicate the existence of the $S^z = \pm 1$ and $S^z = 0$
optical magnons with mass gaps $m_1$ and $m_3$, respectively.
Increasing the spin-orbital coupling leads eventually to an Ising
criticality at $\lambda = \lambda_{c1} = 2 \sqrt{J_{\tau} m/\pi C}$,
where $m_3 = 0$. At $m_3 < 0$ the system occurs in a long-range
ordered N\'eel phase with staggered magnetization  $ \langle S^z _n
\rangle = (-1)^n \zeta (\lambda)$, in which the $Z_2$-symmetry of
model (\ref{spin-anis}) is spontaneously broken. Using the
Ising-model representation (\ref{N-eps}) of the staggered
magnetization of the spin-1 chain,  we find that at $0 < \lambda -
\lambda_{c1} \ll \lambda_{c1}$ the order parameter $\zeta(\lambda)$
follows a power-law increase:
\begin{eqnarray}
    \zeta(\lambda) \sim \left( \frac{\lambda -
    \lambda_{c1}}{\lambda_{c1}}\right)^{1/8}.\label{zeta}
\end{eqnarray}
The transverse spin fluctuations  become incoherent in this phase.
The situation here is entirely similar to that in the spontaneously
dimerized massive phase of a two-chain spin-1/2 ladder\cite{gnt,nt},
where the dimerization kinks make spin fluctuations incoherent.
In the present case, the spontaneously broken $Z_2$ symmetry of the Neel
phase leads to the existence of pairs of massive topological kinks
contributing to a broad continuum with a
threshold at $\omega = m_1 + |m_3|$ (the details of calculation can be found in
Ref.\onlinecite{gnt}):
\bea && \Im
m~\chi^{xx}(q, \omega) \\ &&\quad \sim \frac{1}{\sqrt{m_1|m_3|}}
\frac{\theta(\omega^2 - (q - \pi)^2 v^2 - (m_1 + |m_3|)^2)}
{\sqrt{\omega^2 - (q - \pi)^2 v^2 - (m_1 + |m_3|)^2}}.\nn
\label{chi-incoh} \eea

\medskip

In the N\'eel phase, the orbital sector acquires quantum dynamics
because antiferromagnetic ordering of the spins generates an
effective transverse magnetic field which transforms the classical
Ising model $H_{\tau}$ to a quantum Ising chain. At $\lambda >
\lambda_{c1}$ the spin-orbit term takes the form \be H_{S\tau} = - h
\sum_n (-1)^n \tau^x _n + H'_{S\tau}, \label{stau-new} \ee where $ h
= \lambda \zeta(\lambda) $ and $H'_{S\tau} = - \lambda \sum_n \left(
S^z _n - \la S^z _n \ra \right)\tau^x _n$ accounts for fluctuations.
Since both the orbital and spin sectors are gapped, the main effect
of this term is a renormalization of the mass gaps and group
velocities. The transverse field $h$ gives rise to quantum
fluctuations which decrease the classical value of $\eta^z$ and, at
the same time, lead to a staggered ordering of the orbital
pseudospins in the transverse direction.
Since the orbital sector has a finite susceptibility with respect
to a transverse staggered field,
in the right vicinity of
the critical point $\eta^x$ follows the same power-law increase as
$\zeta$ but with a smaller amplitude: \be \eta^x  \sim \left(
\frac{h}{J_{\tau}} \right) \sim
\sqrt{\frac{\Delta_S}{J_{\tau}}}\left( \frac{\lambda -
\lambda_{c1}}{\lambda_{c1}} \right)^{1/8}.\label{tau_x} \ee This
result is in a good agreement with  previously obtained numerical
results for order parameters (See Fig.~4(a) in
Ref.~\onlinecite{chern2}).

Performing an inhomogeneous $\pi$-rotation of the pseudospins around
the $y$-axis,
$\tau^{x,z}_n \to (-1)^n \tau^{x,z}_n$, $~\tau^y _n \to \tau^y _n$,
we find that at $\lambda > \lambda_{c1}$
the effective  model in the orbital sector
reduces to a ferromagnetic Ising chain in a uniform transverse (pseudo)magnetic field:
\be
H_{\tau; {\rm eff}} = - J_{\tau} \sum_n \tau^z _n \tau^z _{n+1}
- h \sum _n \tau^x _n. \label{QIC}
\ee
Notice that the restriction $\lambda \ll J_{\tau}$, which was imposed in the derivation of the
effective Hamiltonian in the spin sector, now can be released because the spin sector
is assumed to be in the N\'eel phase.
\medskip

At  $h = J_{\tau}$, i.e. at $\lambda = \lambda_{c2}$ where $\lambda_{c2}$ satisfies
the equation
\be
\lambda_{c2} ~\zeta(\lambda_{c2}) = J_{\tau}, \label{orb_tran_point}
\ee
the model (\ref{QIC}) undergoes a 2D Ising transition \cite{kogut,gnt}
to a massive disordered
phase with $\la \tau^z _n \ra = 0$. This quantum critical point can be reached when
$\lambda$ is further increased in the region $\lambda > \lambda_{c1}$.
It is clear from (\ref{orb_tran_point}) that
$\lambda_{c2}$ is of the order of or greater than $J_{\tau}$.
It is reasonable to assume that for such values of $\lambda$
the N\'eel magnetization is close to its nominal value, $\zeta
\sim 1$, implying that  $\lambda_{c2} \sim J_{\tau}$.
 We see that the two Ising transitions are well separated:
\be
\lambda_{c2}/\lambda_{c1} \sim
(J_{\tau}/\Delta_S)^{1/2} \gg 1.\label{isings_separa}
\ee

Thus, in the limit $J_{\tau} \gg \Delta_S$,  the ground-state phase
diagram of the model~(\ref{ham}) consists of three gapped phases
separated by two Ising criticalities, one in the spin sector
($\lambda = \lambda_{c1}$) and the other in the orbital sector
($\lambda = \lambda_{c2}$).
 At $0<\lambda < \lambda_{c1}$ the spin sector
represents an anisotropic spin-liquid while in the orbital sector there is a N\'eel-like ordering of
the pseudospins:
$(-1)^n \la \tau^z _n \ra \equiv \eta^z (\lambda)\neq 0$. At
$\lambda_{c1} < \lambda < \lambda_{c2}$ the orbital degrees of
freedom reveal their quantum nature: the onset of the spin N\'eel
order ($\zeta \neq 0$) is accompanied by the emergence of the
\emph{transverse} component of the staggered pseudospin density:
$(-1)^n \la \tau^x _n \ra \equiv \eta^x (\lambda) \neq 0$. Upon
increasing $\lambda$, the staggered orbital order parameter $\veta$
undergoes a continuous rotation from the $z$-direction to
$x$-direction. At $\lambda = \lambda_{c2}$ a quantum Ising
transition takes place in the orbital sector where $\eta^z$
vanishes. At $\lambda > \lambda_{c2}$ both sectors are long-range
ordered, with order parameters $\zeta, \eta^x \neq 0$. The
dependence of order parameters on $\lambda$ is schematically shown
in Fig.~2(a); this picture is in full qualitative agreement with the
results of the recent numerical studies. \cite{chern2}

\medskip

The crossover between the small and large $\lambda$ limits studied
in this section corresponds to path 1 on the phase diagram shown in
Fig.~\ref{merging}. The path is located in the region $J_{\tau} \gg
\Delta_S$. Starting from the massive phase I and moving along this
path we first observe the \emph{spin-Ising} transition (I $\to$ II)
to the N\'eel phase. Long-range ordering of the spins induces
quantum reconstruction of the initialy classical orbital sector
(i.e. generation of a nonzero $\eta^x$). The \emph{orbital-Ising}
transition (II $\to$ III) takes place inside the spin N\'eel phase.
Of course, feedback effects (that is, orbit affecting spin) become
inreasingly important upon deviating from the critical curve
$\Delta_S J_{\tau} \sim 1$ into phases II and III, especially in the
vicinity of the orbital transition where the spin-orbit coupling is
very strong, $\lambda \sim J_{\tau}$. In this region the behavior of
the spin degrees of freedom is not expected to follow that of an
isolated anisotropic spin-1 chain in the N\'eel phase since the
effect of an ``explicit'' staggered magnetic field $\sim \lambda
\eta^x$ becomes important. We will see a pattern of such behavior in
the opposite limit of ``heavy'' spins, which is discussed in the
next section.


\begin{figure}
\includegraphics[width=0.98\columnwidth]{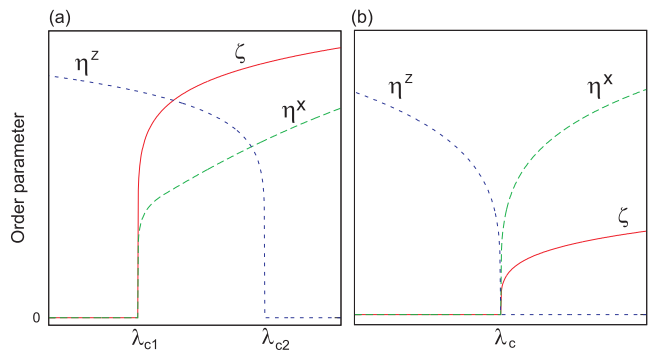}
\caption{\label{fig:op} Schematic diagram of order parameters as
functions of the SO coupling constant $\lambda$. (a) Two Ising
transitions in the $J_{\tau} \gg \Delta_S$ limit. (b) A single
Gaussian transition in the $\Delta_S \gg J_{\tau}$ limit. These two
scenarios correspond to path-1 and path-2 in the phase diagram
(Fig.~1), respectively.}
\end{figure}

\section{Gaussian criticality at ${\bf J}_{\vtau} \vll \vDelta_{\bf S}$}\label{gauss}

In this section we turn to the opposite limiting case: $\Delta_S \gg
J_{\tau}$. Now the spin degrees of freedom constitute the ``fast''
subsystem and can be integrated out to generate an effective action
in the orbital sector. We will show that, in this regime, the
intermediate massive phase where the orbital order parameter $\veta$
undergoes a continuous rotation from $\veta = (0,0,\eta^z)$ to
$\veta = (\eta^x, 0,0)$ no longer exists. Going along path 2,
Fig.~\ref{merging}, which is located in the region $\Delta_S \gg
J_{\tau}$, we find that the two massive phases, I and III, are
separated by a single Gaussian critical line characterized by
central charge $c=1$. On this line the vector $\veta$ vanishes, the
orbital degrees of freedom become gapless
 and represent a spinless Tomonaga-Luttinger liquid characterized by power-law orbital correlations.
\medskip

At $\lambda = 0$ the spin-1 subsystem  represents a disordered,
isotropic spin liquid. Therefore the first nonzero correction to the
low-energy effective action in the orbital sector appears in the
second order in $\lambda$: \bea &&\Delta S^{(2)}_{\tau} = -
\frac{1}{6} \la S^2 _{S\tau} \ra_{S} \\ &&=  - \frac{1}{2} \lambda^2
\sum_{nm} \int  \rd \tau_1 \int  \rd \tau_2~ \la {\bf S}_n (\tau_1)
{\bf S}_{m} (\tau_2)\ra_S ~  \tau^x _{n}(\tau_1) \tau^x
_{m}(\tau_2), \nn \label{sec-ord1} \eea
where $\la \cdots \ra_S$ means averaging over the massive spin degrees of freedom.
According to the decomposition of the spin density, Eq. (\ref{spin-den-ladder}),
the correlation function in (\ref{sec-ord1}) has the structure:
\be
 \la  {\bf S}_l (\tau) {\bf S}_{0} (0)\ra
= (-1)^l f_1 (r/\xi_S) +  f_2 (r/\xi_S). \label{spin_corr_gen}
\ee
Here $\xi_s = v_s / \Delta_S$ is the spin correlation length and ${\bf r} =
(v_s \tau, x)$ is the Euclidian two-dimensional radius-vector.
 $f_1$ and $f_2$ are smooth functions with
the following asymptotic behaviour \cite{gnt} \bea f_1 (x) = C_1
x^{-1/2} e^{-x}, ~~f_2 (x) = C_2 x^{-1} e^{-2x} ~~(x \gg 1),\quad
\label{f-asympt} \eea where $C_1$ and $C_2$ are nonuniversal
constants. DMRG calculations show \cite{sorensen} that $C_2 \ll
C_1$; for this reason the contribution of the smooth part of the
spin correlation function can be neglected in (\ref{sec-ord1}).
\medskip

Integrating over the relative time $\tau_- = \tau_1 - \tau_2$ we
find that the spin-orbit coupling generates a pseudospin
$xx$-exchange with the following structure: \be H'_{\tau} = \sum_n
\sum_{l \geq 1} (-1)^{l+1} J' _{\tau} (l) \tau^x _n \tau^x _{n+l}
\label{extra} \ee Here the exchange couplings exponentially decay
with the separation $l$, $J' _{\tau} (l) \sim (\lambda^2 / \Delta_S)
\exp (- la_0 / \xi_S)$, so the summation in (\ref{extra}) actually
extends up to $l \sim \xi_S /a_0$. In the Heisenberg model $\xi_S$
is of the order of a few lattice spacings, so for a qualitative
understanding it would be sufficient to consider the $l=1$ term as
the leading one and treat the $l=2$ term as a correction. Making a
$\pi/2$ rotation in the pseudospin space, $\tau^z _n \to \tau^y _n$,
$\tau^y _n \to -\tau^z _n$,  we pass to the conventional notations
and write down the effective Hamiltonian for the orbital degrees of
freedom as a perturbed XY spin-1/2 chain: \be H^{\rm eff}_{\tau} =
\sum_n \left( J_{x} \tau^x _{n} \tau^x _{n+1} + J_y \tau^y _{n}
\tau^y _{n+1}\right) + H'_{\tau} . \label{xy-chain} \ee where \be
 H'_{\tau} = - J' _x \sum_n \tau^x _n \tau^x _{n+2} + \cdots .\label{xx-pert}
\ee
Here $J_y = J_{\tau}$, $J_x = J'_{\tau} (1) > 0$ and $J'_x = J' _{\tau} (2) > 0$.
By order of magnitude
$J'_x < J_x \sim \lambda^2 / \Delta_S$.
\medskip

In the absence of the perturbation $H' _{\tau}$, the model
(\ref{xy-chain}) represents a spin-1/2 XY chain which for any
nonzero anisotropy in the basal plane ($J_x \neq J_y$) has a N\'eel
long-range order in the ground state and a massive excitation
spectrum. This follows from the Jordan-Wigner transformation \be
\tau^z _n = 2 a^{\dagger}_n a_n - 1, ~~~ \tau^{+} _n = \tau^x _n+
\ri \tau^y _n = 2 a^{\dagger}_n e^{i \pi \sum_{j<n} a^{\dagger}_j
a_j} \label{JW} \ee which maps the XY chain onto a model of complex
spinless fermions with a Cooper pairing:\cite{lsm} \bea H^{\rm
eff}_{\tau} &=& (J_x + J_y) \sum_{n} \left( a^{\dagger}_n a_{n+1} +
h.c. \right)\nn \\ &+& (J_x - J_y) \sum_{n} \left( a^{\dagger}_n
a^{\dagger}_{n+1} + h.c. \right). \label{spinless_ferm} \eea By
increasing $\lambda$ (equivalently, decreasing $J_{\tau}$) the model
(\ref{spinless_ferm}) can be driven to a XX quantum critical point,
$J_x =J_y(1)$, i.e. $\lambda = \lambda_c \sim \sqrt{J_{\tau}
\Delta_S}$, where the the system acquires a continuous U(1)
symmetry. At this point the Jordan-Wigner fermions become massless
and the system undergoes a continuous quantum transition.
\medskip

The transition
is associated with reorientation of the pseudospins.
Away from the Gaussian criticality the effective orbital Hamiltoian is invariant under
$Z_2 \times Z_2$ transformations: $\tau^{x} _n \to -\tau^{x} _n$,
 $\tau^{z} _n \to -\tau^{z} _n$.
 In massive phases
this symmetry is spontaneously broken.
Making a back rotation from $\tau^y$ to $\tau^z$ we conclude that
at $J_y  > J_x$ ($\lambda < \lambda_c$)
$ \eta^z \neq 0$, $\eta^x = 0$, while at $J_y  < J_x$
($\lambda > \lambda_c$)
$\eta^z = 0$, $\eta^x \neq 0$.
Both $\eta^z$ and $\eta^x$ vanish at the critical point, so contrary to the
case $J_{\tau} \gg \Delta_S$, here there is no region of
their coexistence.

\medskip

The passage to the continuum limit for the model (\ref{xy-chain})
based on Abelian bosonization is discussed in Appendix \ref{XY}.
There we show that the perturbation $H'_{\tau}$ adds a marginal
four-fermion interaction $g = J'_x (2)/\pi v \ll 1$ to the
free-fermion model (\ref{XY-cont}). In the spin-chain language,
this is equivalent to adding a weak ferromagnetic $zz$-coupling.
In the limit of weak XY anisotropy, $|\lambda -
\lambda_c|/\lambda_c \ll 1$, the low-energy properties of the
orbital sector are described by a
quantum sine-Gordon model (all notations are explained in Appendix
\ref{XY}) \bea {\cal H} = \frac{u}{2} \left[ K \Pi^2 + \frac{1}{K}
\left( \p_x \Phi \right)^2  \right] + \frac{2 \gamma}{\pi \alpha}
\cos \sqrt{4\pi} \Theta, \label{LL-SG1} \eea where \be \gamma \sim
J_{\tau} \left( \frac{\lambda - \lambda_c}{\lambda_c} \right), ~~~K
= 1 + 2g + O(g^2). \label{gamma-crit} \ee The U(1) criticality is
reached at $\lambda = \lambda_c$ where, due to a finite value of
$g$, the orbital degrees of freedom represent a Tomonaga-Luttinger
liquid. Close to the criticality, the spectral gap in the orbital sector
scales as the
renormalized
mass of the sine-Gordon model (\ref{LL-SG1}):
\be
M_{\rm orb} \sim \Big|  \frac{\lambda - \lambda_c}{\lambda_c}
\Big|^{\frac{K}{2K-1}}. \label{orb_gap}
\ee

Strongly fluctuating physical fields acquire coupling
dependent scaling dimensions. In particular, according to the
bosonization rules,\cite{gnt} the staggered pseudospin densities are
expressed in terms of the vertex operators, \bea (-1)^n \tau^x _n
&\equiv& n^x (x) \sim \sin \sqrt{\pi} \Theta(x),\nn\\ (-1)^n \tau^z
_n &\equiv& n^z (x) \sim \cos \sqrt{\pi} \Theta(x),
\label{stag_pseudospin} \eea both with scaling dimension $d=1/4K$.
This anomalous dimension
 determines the power-law behaviour of the average staggered densities
close to the criticality:
\bea
\eta^z (\lambda) \sim (\lambda_c - \lambda)^{1/4K}, && \lambda < \lambda_c \nn\\
\eta^x (\lambda) \sim (\lambda - \lambda_c)^{1/4K}, && \lambda > \lambda_c.
\label{order_gauss}
\eea

A finite staggered pseudospin magnetization $\eta^x$ at $\lambda >
\lambda_c$ generates an effective \emph{external} staggered magnetic
field in the spin sector: \be H_S \to \bar{H} = H_S + H'_S, ~~~H'_S
= - h_S \sum_n (-1)^n S^z _n, \label{stag-field-spin} \ee where $h_S
= - \lambda \eta^x$. The spectrum of the Hamiltonian $\bar{H}$ is
always massive. This can be easily understood within the Majorana
model (\ref{Maj-Tsv}). According to (\ref{N-eps}), in the continuum
limit, the sign-alternating component of the spin magnetization,
$N^3 \sim (-1)^n S^z _n$, can be expressed in terms of the order and
disorder fields of the degenerate  triplet of 2D disordered Ising
models: $N^3 \sim \mu_1 \mu_2 \s_3$. In the leading order, the
magnetic interaction $H'_S$ gives rise to an effective magnetic
field $h_3 = h_S \la \mu_1 \mu_2 \ra$ applied to the third Ising
system: $h_3 \s_3$. The latter always stays off-critical.
\medskip

Since in the Haldane phase the spin correlations are short-ranged,
close to the transition point the induced staggered magnetization
$\zeta$ can be estimated using linear response theory. Therefore, at
$0 < \lambda - \lambda_c \ll \lambda_c$, $\zeta$ follows the same
power-law increase as that of $\eta^x$ but with a smaller amplitude:
\be \zeta \sim \frac{h_S}{\Delta_S} \sim
\left(\frac{J_{\tau}}{\Delta_S} \right)^{1/2} \left(\frac{\lambda -
\lambda_c}{\lambda_c} \right)^{1/4K} \label{induced_zeta} \ee So, in
the part of the phase C, Fig. \ref{merging}, where $\Delta_S\gg
J_{\tau}$, the $\eta^x$-orbital order, being the result of a
spontaneous breakdown of a $Z_2$ symmetry $\tau^x _n \to - \tau^x
_n$, acts as an effective staggered magnetic field applied to the
spins and \emph{induces} their N\'eel alignment. This fact is
reflected in a coupling dependent, nonuniversal exponent $1/4K$
characterizing the increase of the staggered magnetization at
$\lambda > \lambda_c$. The order parameters as functions of
$\lambda$ in the $\Delta_S \gg J_{\tau}$ limit is schematically
shown in Fig.~2(b).
\medskip

As already mentioned, the absence of a small parameter in the regime
of strong hybridization, $J_{\tau} \sim J_S \sim \lambda$, makes the analysis of
the phase diagram in this region not easily accessible by analytical tools.
Nevertheless some plausible arguments can be put forward to comment on the topology
of the phase diagram. It is tempting to treat the curve $J_{\tau} \Delta_S /\lambda^2\sim 1$
as a single critical line going throughout the whole phase plane ($J_{\tau}/\lambda, \Delta_S/\lambda$).
If so, we then can expect that there exists a special singular point located in the region
$J_{\tau} \Delta_S /\lambda^2\sim 1$.
This expectation is based on the
fact that at $J_{\tau} \gg \Delta_s$ limit the transition is of the Ising type
and the spontaneous spin magnetization below the critical curve follows
the law $\zeta \sim (\lambda - \lambda_{c1})^{1/8}$ with a \emph{universal} critical exponent,
whereas at $J_{\tau} \ll \Delta_s$ the spin magnetization has a different,
\emph{nonuniversal} exponent, $\zeta \sim (\lambda - \lambda_c)^{1/4K}$.
Continuity considerations make it
very appealing
to suggest that at the special point the Tomonaga-Luttinger liquid parameter takes
the value $K=2$, and the two power laws match. Since the central charges of two Ising
and one Gaussian criticalities satisfy the relation $1/2 + 1/2 = 1$, the singular
point must be a point where the two Ising critical curves merge into a single Gaussian one.


\section{Dynamical sin susceptibility and NMR relaxation rate in the vicinity of Gaussian criticality}
\label{V}

It may seem at the first sight that, in the regime $\Delta_S \gg J$,
the spin degrees of freedom
which have been integrated out remain massive across the orbital
Gaussian transition, and the spectral weight of the staggered spin
fluctuations
is only nonzero in the high-energy region $\omega \sim \Delta_S$.
However, this conclusion is only correct for the zeroth-order
definition of the spin field ${\bf N}_0 (x)$, given by Eq.
(\ref{N-eps}), with respect to the spin-orbit interaction. In fact,
the staggered magnetization hybridizes with low-energy orbital modes
via SO coupling already in the first order in $\lambda$ and thus acquires
a low-energy projection which contributes to a nonzero spectral weight displayed by
the dynamical spin  susceptibility at energies well below the Haldane gap.
\medskip

To find the low-energy projection of the field $N^z ({\bf r})$, we must fuse
the local operator $N^z _0 ({\bf r})$ with the perturbative part of the total action.
Keeping in mind that close to and at the Gaussian criticality most strongly fluctuating
fields are the staggered components of the orbital polarization, we approximate the SO
part of the Euclidian action by the expression
\be
 S_{S\tau} \simeq \frac{\lambda a_0}{v_S} \int \rd^2 {\bf r}~N^z ({\bf r})
 n^x ({\bf r}),
 \ee
where ${\bf r} = (v_S \tau, x)$ is the two-dimensional radius vector (here $\tau$ is the
imaginary time). We thus construct
\bea
&& N^z_P ({\bf r}) =  \langle e^{-S_{S\tau}} N^z({\bf r})  \rangle \nn\\
&& = N_0 ^z ({\bf r}) - \frac{\lambda a_0}{v_S} \int \rd^2
{\bf r}_1 \la N_0 ^z ({\bf r})N_0 ^z ({\bf r}_1) \ra_S ~n^x ({\bf
r}_1) \nn\\
&& ~+~ O(\lambda^2), \label{N_z-proj} \eea where averaging is
done over the unperturbed, high-energy spin modes. For simplicity,
here we neglect the anisotropy of the spin-liquid phase of the S=1
chain and  use formula (\ref{NN}). The spin correlation function is
short-ranged. Treating the spin correlation length $\xi_S \sim
v_S/\Delta_S$ as a new lattice constant (new ultraviolet cutoff) and
being interested in the infrared asymptotics $|{\bf r}| \gg \xi_S$,
we can replace in (\ref{N_z-proj}) $n^x ({\bf r}_1)$ by $n^x ({\bf
r})$. The integral \bea && \int \rd^2\vrho~ \la N_0^z (\vrho)N_0^z
({\bf 0})\ra_S \\ &&\quad \sim  \frac{1}{a^2 _0} (a/\xi_S)^{3/4}
\int_0 ^{\infty} d\rho~ \rho \sqrt{\xi_s/\rho}~ e^{-\rho/\xi_S} \sim
(\xi_S / a_0)^{5/4}.\nn \eea So the first-order low-energy
projection of the staggered magnetization is proportional to \be
N^z_P ({\bf r})\sim \frac{\lambda}{\Delta_S} \left(
\frac{\xi_S}{a_0} \right)^{1/4} n^x ({\bf r}). \label{proj_fin} \ee
This result clarifies the essence of the hybridization effect:
close to the Gaussian criticality the spin fluctuations
acquire a finite spectral weight in the low-energy region,
$\omega \ll \Delta_S$, 
$q \sim \pi$,
which is contributed by
orbital fluctuations and can be probed in magnetic inelastic neutron scattering
experiments and NMR measurements.
\medskip

Away from but close to the Gaussian criticality the behavior
of the dynamical spin susceptibility $\Im m \chi(q, \omega)$ is determined by the excitation
spectrum of the sine-Gordon model for the dual field, Eq.(\ref{LL-SG1}). Since $K > 1$, it
consists of kinks, antikinks carrying the mass
$M_{\rm orb}$, and their bound states (breathers) with masses (see e.g. Ref. \onlinecite{gnt})
\bea
&& M_j = 2M_{\rm orb} \sin (\pi j/2\nu),\nn\\
&& j=1,2, \ldots \nu - 1, ~~ \nu = 2K-1
\eea
Since $K-1 = 2g$ is small, there will be only the first breather in the spectrum,
with mass $M_1 = 2M_{\rm orb} (1 - 2\pi^2 g^2)$.
The sine-Gordon model is integrable, and the asymptotics of its correlation functions
in the massive regime have been calculated using the form-factor
approach (see for a recent review \onlinecite{essler}).
Here we utilize some of the known results.
At $\lambda < \lambda_c$ the operator $n^x \sim \sin \sqrt{\pi} \Theta$
has a nonzero matrix element between the vacuum  and the first breather
state. This form-factor contributes to a coherent peak in the dynamical spin susceptibility
at frequencies much smaller than than the Haldane gap:
\bea
\Im m \chi(q, \omega, T=0) &=& A (\lambda/\Delta_S)^2
\delta [\omega^2 - (q-\pi)^2 v^2 - M^2 _1] \nn\\
&+& \Im m \chi_{\rm cont}(q, \omega, T=0).
\label{breather}
\eea
Here $A$ is a constant and the second term is the contribution of a
multi-kink continuum of states with a threshold at $\omega = 2M_{\rm orb}$.
At $\lambda > \lambda_c$ the spectral properties of the operator
$\cos \sqrt{\pi} \Theta$ coincide with those of the operator
$\sin \sqrt{\pi} \Theta$ at $\lambda < \lambda_c$. For symmetry reasons \cite{essler},
this operator does not couple to the first breather, so that at
$\lambda > \lambda_c$ $\Im m \chi(q, \omega)$ will only display the kink-antikink
scattering continuum.
\medskip

We see that, due to spin-orbit hybridization effects, the spin sector
of our model loses the properties of a spin liquid already in a noncritical
orbital regime. This tendency gets strongly enhanced at the orbital Gaussian
criticality
($M_{\rm orb} \to 0$) where all multi-particle processes
merge, and
the spin correlation function exhibits
an algebraically decaying asymptotics
\be
 \la N^z ({\bf r})N^z ({\bf 0})\ra \simeq \la N^z _P ({\bf r})N^z _P ({\bf 0})\ra
 \sim \left(\frac{\lambda}{\Delta_S} \right)^2 \left( \frac{a}{r} \right)^{\frac{1}{2K}},
 \ee
implying that the spin sector of the model becomes
reminiscent of Tomonaga-Luttinger liquid. In this limit (here for simplicity
we consider the $T=0$ case)
the dynamical spin susceptibility is given by\cite{schulz}
\be
\Im m \chi(q, \omega, T=0) \sim (\lambda/\Delta_S)^2\left[\omega^2  - v^2 (q - \pi)^2 \right]^{\frac{1}{4K} - 1}. \label{chi_spinorbit}
\ee

The NMR relaxation rate probes the spectrum of local spin fluctuations
\[
\
\frac{1}{T_1} = A^2 T \lim_{\omega \to 0} \frac{1}{\omega}\sum_q
\Im m \chi^{zz} (q, \omega, T) \label{rela_rate}
\]
where $A$ is an effective hyperfine constant. In spin-liquid regime
of an isolated spin-1 chain, the existence of a Haldane gap makes
$1/T_1$ exponentially suppressed\cite{sagi}: $1/T_1 \sim
\exp(-2\Delta_S/T)$. The admixture of low-energy orbital states in
the spin-fluctuation spectrum drastically changed this result. A
simple power counting argument\cite{giam} leads to a power-law
temperature dependence of the NMR relaxation rate: \be \frac{1}{T_1}
\sim A^2 \left(  \frac{\lambda}{\Delta_s}\right)^2 T^{\frac{1}{2K} -
1} \label{final_NMR} \ee This result is valid not only exactly at
the Gaussian criticality but also in its vicinity provided that the
temperature is larger than the orbital mass gap. By construction
(see the preceding section) $K \geq 1$. This means that the exponent
$1/2K - 1$ is \emph{negative} and the NMR relaxation rate
\emph{increases} on lowering the temperature. It is worth noticing
that such regimes are not unusual for Tomonaga-Luttinger phases of
frustrated spin-1/2 ladders.\cite{sato} For our model, such behavior
of $1/T_1$ would be a strong indication of an extremely quantum
nature of the collective orbital excitations.\cite{footnote}


\section{Behavior in a magnetic field: Quantum Ising transition in orbital sector}
\label{mag_field}

We have seen in Sec.\ref{heavy_orbit} that, due to spin-orbit coupling,
the N\'eel ordering of the spins is
accompanied by the emergence of
quantum effects in the orbital sector: the classical orbital Ising chain transforms
to a quantum one. In this section we briefly comment on
a similar situation that can arise upon
application of a uniform external magnetic field $h$.
\medskip

Since the spin-1 chain is massive, it will acquire a finite ground-state magnetization
$\la S^z \ra$ only when the magnetic field, $h$, is higher than the critical
value $h_{c1} \sim \Delta_S$, corresponding to the
commensurate-incommensurate (C-IC) transition. According to the definition
(\ref{vector_currents}), a uniform magnetic field along the $z$-axis,
${\cal H}_{\rm mag} = - h I^z$, mixes up
a pair of Majorana fields, $\xi^1$ and $\xi^2$, and splits the spectrum
of $S^z = \pm 1$ excitations (the $S^z=0$ modes are unaffected by the field).
At $h = h_{c1}$ the gap in the spectrum of the $S^z=1$ excitations closes, and
at $h>h_{c1}$ these modes condense giving rise to a finite magnetization.
Once
$\la S^z \ra \neq 0$,
the effective Hamiltonian of the $\tau$-chain becomes
\be
\bar{H}_{\tau} = J_{\tau} \sum_n \tau^z _{n} \tau^z _{n+1} -
\Delta_{\tau}\sum_n  \tau^x _{n}, ~~~~\Delta_{\tau} = \lambda \la S^z \ra.
\label{eff-tau-chain}
\ee
Here we ignore the
fluctuation term
that couples
$\tau^x _n$ to
$\Delta S^z _n = S^z _n - \la S^z _n \ra $.
\medskip


One should keep in mind that there exists the second C-IC transition at a higher field $h_{c2}$
associated with full polarization
of the spin-1 chain.
To simplify further analysis, let us assume that the range of magnetic fields
$h_{c1} < h < h_{c2}$, where an isolated spin-1 chain has an incommensurate,
gapless ground state, is sufficiently broad.
This can be easily achieved in
the biquadratic model (\ref{biquad-chain})
with $\beta \sim 1$, in which case the Haldane gap -- and hence $h_{c1}$ --
is small, and the effects associated with the second C-IC transition can be neglected.

\medskip

Now, by increasing the magnetic field $h$ in the region $h>h_{c1}$,
 the effective orbital chain (\ref{eff-tau-chain}) can be driven
to an Ising criticality. The induced transverse ``magnetic field'' $\Delta_{\tau}$
is proportional to a nonzero magnetization of the spin-1 chain. If $\lambda/J_{\tau}$
is large enough, then upon increasing the field
the effective quantum Ising chain (\ref{eff-tau-chain}) can reach the point
$\Delta_{\tau} (h^*) = J_{\tau}$ where the Ising transition occurs.
This will happen at some field $h = h^* > h_{c1}$.
In the region $|h-h^*|/h^* \ll 1$ the quantum Ising $\tau$-chain
will be slightly off-critical. Due to the SO coupling, these massive orbital excitations
will interact with the gapless $S^z= \pm 1$ spin modes. However, this interaction
can only give rise
to the orbital mass renormalization (i.e. a small shift
of the Ising critical point) and a group velocity renormalization of the
spin-doublet modes. For this reason
we do not expect the aforementioned spin-orbital fluctuation term
to cause any qualitative changes.
\medskip

The above discussion reveals an interesting fact:  a sufficiently strong magnetic field
acting on the spin degrees of freedom
can affect the orbital structure of the chain and drive it to a quantum Ising transition.
The difference with the situation discussed in Sec.\ref{heavy_orbit}
is that the external magnetic field induces a uniform spin polarization which, in turn,
gives rise to a uniform transverse orbital ordering $\la  \tau^x _n\ra \neq 0$.
Thus, the classical long-range orbital order $\la \tau^z _n \ra = (-1)^n \eta^z$,
present at $h < h^*$, disappears in the region  $h > h^*$, where the orbital degrees of freedom
are characterized by a transverse ferromagnetic polarization,  $\la \tau^x \ra \neq 0$.


\section{Conclusion and discussion}\label{con}

In this paper, we have proposed and analyzed a 1D spin-orbital model
in which a spin-1 Haldane chain is locally coupled to an orbital
Ising chain by an on-site term $\lambda\,\tau^x S^z$ originating
from relativistic spin-orbit (SO) interaction. The SO term not only
introduces anisotropy to the spin sector, but also gives quantum
dynamics to the orbital degrees of freedom. We approach this problem
from well defined limits where either the spin or the orbital sector
is strongly gapped and becomes a `fast' subsystem which can be
integrated out. By analyzing the resultant effective action of the
remaining `slow' degrees of freedom, we have identified the stable
massive and critical phases of the model which are summarized in a
schematic phase diagram shown in Fig.~1.

In the limit dominated by a large orbital gap, i.e.  $J_{\tau} \gg
\Delta_S$, integrating out the orbital variables gives rise to an
easy-axis spin anisotropy $D (S^z)^2$ where $D \sim
-\lambda^2/J_{\tau}$. As $\lambda$ increases, the disordered Haldane
spin liquid undergoes an Ising transition into a magnetically
ordered N\'eel state. The presence of antiferromagnetic spin order
$\zeta$ in the N\'eel phase in turn generates an effective
transverse field $h \sim \lambda \zeta$ acting on the orbital Ising
variables. The orbital sector which is described by the Hamiltonian
of a quantum Ising chain reaches criticality when $h = J_{\tau}$. In
between the two Ising critical points lies an intermediate phase
(phase II in Fig.~1) where both Ising order parameters $\eta^x$ and
$\eta^z$ are nonzero. Such a two-stage ordering scenario illustrated
by path 1 in the phase diagram (Fig.~1) has been confirmed
numerically by recent DMRG calculations. \cite{chern2}
Interestingly, the orbital Ising transition can also be induced by
applying a magnetic field to the spin sector. As the field strength
is greater than the Haldane gap, a field-induced magnon condensation
results in a finite magnetization density $\langle S^z \rangle$ in
the linear chain. Thanks to the SO coupling, the orbital sector
again acquires a transverse field $h \sim \lambda \langle
S^z\rangle$ and becomes critical when $h = J_{\tau}$.

A distinct scenario of the orbital reorientation transition $\eta^z
\to \eta^x$ occurs in the opposite limit $\Delta_S \gg J_{\tau}$.
This time we integrate out the fast spin subsystem and obtain a
perturbed spin-1/2 XY Hamiltonian for the orbital sector. The
effective exchange constants are given by $J_x \sim
\lambda^2/\Delta_S$ and $J_y = J_{\tau}$. As $\lambda$ is varied,
the orbital sector reaches a Gaussian critical point when $J_x =
J_y$, at which the system acquires an emergent U(1) symmetry. The
orbital order parameter goes directly from $\veta = (0, 0,\eta^z)$
to $(\eta^x, 0, 0)$ in this single-transition scenario (illustrated
by path 2 in Fig.~1). Both order parameters $\eta^x$ and $\eta^z$
vanish at the critical point. We have shown that spin-orbital hybridization effects
near the
Gaussian transition  lead to the appearance of a non-zero
spectral weight of the staggered spin density well below the Haldane gap
-- the effect which can be detected by
inelastic neutron scattering experiments and NMR measurements.

The stability analysis of the orbital Gaussian criticality in the original
lattice model (\ref{ham}), done in Appendix \ref{XY}, has shown that this
critical regime is protected by the $\tau^z \to - \tau^z$ symmetry of
the underlying microscopic model. This symmetry will be broken in the
presence of an orbital field $\delta \sum_n \tau^z _n$ which removes degeneracy
between the
local orbitals $d_{zx}$ and $d_{yz}$ and adds a "magnetic"
field along the $y$-axis in the effective XY model (\ref{xy-chain}).
Such perturbation will drive the orbital sector away from the Gaussian criticality.
The same argument applies to a perturbation with the structure $\beta \sum_n S^z _n\tau^z _n$
which also breaks the aforementioned symmetry.
Integrating over the spins will generate an extra term $\sim \lambda \beta
\sum_n (\tau^x _n \tau^y _{n+1} + \tau^y _n \tau^z _{n+1})$
which, in the continuum limit, translates to $\lambda \beta \sin \sqrt{4\pi} \Theta$.
As explained in Appendix \ref{XY}, such perturbation will keep the orbital sector
gapped with coexisting $\eta^x$ and $\eta^z$ orderings.

Since the analysis presented in this paper is done in the limiting
cases, precise predictions on the detailed shape of the phase
diagram or on the behavior of correlation functions in the regime of
strong hybridization of spin and orbital degrees of freedom, where
all interactions included in the model are of the same order, are
beyond our reach and require further numerical calculations. On the
other hand, the continuity and scaling analysis allow us to believe
that the global topology of the phase diagram and character of
critical lines  are given correctly. Finally the spin-orbital model
Eq.~(\ref{ham}) can be generalized to the zigzag geometrical where
two parallel spin-1 chains are coupled to a zigzag Ising orbital
chain via on-site SO interaction. The zigzag case is closely related
to the quasi-1D compound CaV$_2$O$_4$. While the
two-Ising-transitions scenario is expected to hold in the $J_{\tau}
\gg \Delta_S$ regime, the counterpart of Gaussian criticality in the
zigzag chain remains to be explored and will be left for future
study.

\section*{Acknowledgements}

The authors are grateful to Andrey Chubukov,  Fabian Essler,
Vladimir Gritsev,
Philippe Lecheminant and  Alexei Tsvelik for stimulating
discussions. A.N. gratefully acknowledges hospitality of the Abdus
Salam International Centre for Theoretical Physics, Trieste, where
part of this work has been done. He is also supported by the grants
GNSF-ST09/4-447 and IZ73Z0-128058/1. G.W.C. acknowledges the support
of ICAM and NSF grant DMR-0844115. N.P. acknowledges the support
from NSF grant DMR-1005932 and ASG "Unconventional magnetism". G.W.C. and N.P. also thank
the hospitality of the visitors program at MPIPKS, where the part of the work on this manuscript has been done.

\appendix


\section{Ising correlation function}\label{ising_corr}

In this Appendix we estimate the correlation function $\Gamma^{xx}_{nm} (\tau)
= \la \tau^x _n (\tau) \tau^x _m (0) \ra$, where the averaging is performed over the
ground state of the Ising Hamiltonian $H_{\tau} = J_{\tau} \sum_{n}\tau^z _n \tau^z _{n+1} $,
and $\tau^x _n (\tau) = e^{\tau H_{\tau}}\tau^x _n  e^{-\tau H_{\tau}}$.
\medskip

It proves useful to make a duality transformation:
\[
\tau^z _n \tau^z _{n+1} = \mu^x _n, ~~~\tau^x _n = \mu^z _n \mu^z _{n+1}.
\]
The new set of Pauli matrices $\mu^a _n$ represents disorder operators.
The Hamiltonian and correlation function become:
\bea
&& H \to  J_{\tau} \sum_{n} \mu^x _n, \label{new_ising_ham}\\
&& \Gamma^{zz}_{nm} (\tau) \to \la  \mu^z _n (\tau) \mu^z _{n+1}
(\tau) \mu^z _m (0)\mu^z _{m+1}(0) \ra. \label{ham_corr_dualrep}
\eea The most important fact about the dual representation is the
additive, single-spin structure of the Hamiltonian: the latter
describes noninteracting spins in an external ``magnetic field''
$J_{\tau}$. Notice that by symmetry $ \la  \mu^z _n \ra = 0. $
Therefore the correlation function in (\ref{ham_corr_dualrep}) has
an ultralocal structure: \be \Gamma^{xx}_{nm} (\tau) = \delta_{nm}
Y^2 (\tau), ~~~ Y(\tau) = \la \mu^z _n (\tau) \mu^z _n (0)  \ra.
\label{corr_fun_struc} \ee The time-dependence of the disorder
operator can be explicitly computed,
\[
\mu^z _n (\tau) = e^{\tau J_{\tau} \mu^x _n} \mu^z _n e^{-\tau J_{\tau} \mu^x _n}
= \mu^z _n \cosh (2J_{\tau} \tau)  - \ri \mu^y _n \sinh  (2J_{\tau} \tau).
\]
Therefore (below we assume that $\tau > \tau'$) \bea Y(\tau - \tau')
&=& \cosh 2J_{\tau} (\tau - \tau') + \la \mu^x \ra \sinh 2J_{\tau}
(\tau - \tau') \nn\\ &=& \exp [- 2|J_{\tau}| (\tau - \tau')].
\label{Y} \eea Here we used the fact that, in the ground state the
Hamiltonian $H_{\tau}$, $\la \mu^x \ra = - {\rm sgn~}J_{\tau}$.
Thus, as expected for the 1D Ising model, the correlation function
$\Gamma^{xx}_{nm} (\tau)$ is local in real space and decays
exponentially with $\tau$: \be \Gamma^{xx}_{nm} (\tau) = \delta_{nm}
\exp \left( - 4J_{\perp}|\tau| \right). \label{corr_func_fin} \ee

\section{Perturbed XY chain, Eq. (\ref{xy-chain})}\label{XY}

In this Appendix we analyze the perturbation (\ref{xx-pert}) to the XY spin chain
(\ref{xy-chain})
and show that at the XX point it
represents a marginal perturbation
which transforms the free-fermion regime to a Gaussian criticality
describing a Luttinger-liquid behavior of the orbital degrees of freedom.
\medskip

Using the Jordan-Wigner transformation (\ref{JW}) we rewrite
(\ref{xx-pert}) as $H'_{\tau}  = H' _1 + H'_2$, where \bea H'_1 &=&
\frac{J' _x (2)}{2}
\sum_n (a^{\dagger}_n a_{n+2} + h.c.)( a^{\dagger}_{n+1} a_{n+1} - \frac{1}{2}),\quad \label{h'1}\\
H'_2 &=& \frac{J' _x (2)}{2} \sum_n (a^{\dagger}_n a^{\dagger}_{n+2}
+ h.c.) ( a^{\dagger}_{n+1} a_{n+1} - \frac{1}{2}).\quad
 \label{h'2}
\eea Assuming that $|J_x - J_y|, J'_x \ll J_x + J_y$, we pass to a
continuum description of the XY chain in terms of chiral, right (R)
and left (L), fermionic fields based on the decomposition (to
simplify notations we set here $a_0 = 1$): $ a_{n} \to (-\ri)^n R
(x) + \ri^n L(x). $ Then the Hamiltonian density of the XY model
takes the form: \bea {\cal H}_{XY} (x) = - \ri v \left( R^{\dagger}
\p_x R - L^{\dagger} \p_x L \right) - 2\ri \gamma \left( R^{\dagger}
L^{\dagger} - h.c. \right),\nn \!\!\\ \label{XY-cont} \eea where
$\gamma = J_x - J_y$. Standard rules of Abelian
bosonization\cite{gnt} transform (\ref{XY-cont}) to a quantum
sine-Gordon model: \be {\cal H}_{XY} (x) = \frac{v}{2} \left[ \Pi^2
+ \left(\p_x \Phi  \right)^2 \right] + \frac{2 \gamma}{\pi \alpha}
\cos \sqrt{4\pi} \Theta, \label{SG} \ee where $v = 2(J_x + J_y)a_0$
is the Fermi velocity, $\Pi (x) = \p_x \Theta (x)$ is the momentum
conjugate to the scalar field $\Phi (x) = \Phi_R (x) + \Phi_L (x)$, and $\Theta(x)
= - \Phi_R (x) + \Phi_L (x)$ is the field dual
to $\Phi (x)$. Here $\Phi_{R,L}(x)$ are chiral components of the scalar field.
Using the fact that the fermions are spinless, one can impose the condition $[\Phi_R (x), \Phi_L (x')] = i/4$ and thus make sure that the bosonization rules correctly reproduce the
anticommutation relations $\{ R(x), L(x') \} = \{ R(x), L^{\dagger}(x') \} = 0$.
An explicit introduction of the so-called Klein factors becomes necessary when
bosonizing
fermions with an internal degree of freedom, such
as spin 1/2, chain index {\it etc}, which is not the case here.

\medskip

Let is find the structure of the perturbation (\ref{xx-pert}) in the
continuum limit. First of all we notice that \bea &&
a^{\dagger}_{n+1} a_{n+1} - 1/2 \equiv ~:a^{\dagger}_{n+1} a_{n+1}:
\nn \\ && \quad \to (:R^{\dagger} R: + :L^{\dagger} L: ) +
(-1)^{n+1} (R^{\dagger} L + L^{\dagger} R)
\nn\\
&&\quad= \frac{1}{\sqrt{\pi}} \p_x \Phi + \frac{(-1)^n}{\pi \alpha}
\sin \sqrt{4\pi} \Phi.  \label{norm_den} \eea Similarly \bea &&
a^{\dagger}_{n} a_{n+2} + h.c. \nn\\ && \quad \to - 2 \left[
(:R^{\dagger} R: + :L^{\dagger} L: ) + (-1)^n
 (R^{\dagger} L + L^{\dagger} R) \right]\nn\\
&&\quad= - 2 \left[  \frac{1}{\sqrt{\pi}} \p_x \Phi
-\frac{(-1)^n}{\pi \alpha} \sin \sqrt{4\pi} \Phi \right].
\label{pert1} \eea Dropping Umklapp processes
$R^{\dagger}(x)R^{\dagger}(x+\alpha) L(x+\alpha)L(x) + h.c. \sim
\cos \sqrt{16 \pi} \Phi $ as strongly irrelevant (with scaling
dimension 4) at the $XX$ criticality and ignoring interaction of the
fermions in the vicinity of the same Fermi point, we find that \bea
( a^{\dagger}_n a_{n+2} + h.c.) (a^{\dagger}_{n+1} a_{n+1} -
1/2)\Big|_{\rm smooth} \nn\\ \to - 8 :R^{\dagger} R: :L^{\dagger} L:
= 2 \left[ \Pi^2 - (\p_x \Phi)^2 \right].\label{first} \eea We see
that the perturbation $H'_1$ generates a marginal four-fermion
interaction to the free-fermion model (\ref{XY-cont}), thus
transforming the model (\ref{xy-chain}) to an XYZ model with a weak
ferromagnetic $(zz)$-coupling. This interaction can be incorporated
into the Gaussian part of the bosonic theory (\ref{SG}) by changing
the compactification radius of the field $\Phi$: \bea {\cal H} &=&
{\cal H}_{XY} + {\cal H}'_1 \nn \\ &=& \frac{u}{2} \left[ K \Pi^2 +
\frac{1}{K} \left( \p_x \Phi \right)^2 \right] - \frac{2 \gamma}{\pi
\alpha} \cos \sqrt{4\pi} \Theta. \label{LL-SG} \eea Here $u$ is the
renormalized velocity and $K$ is the interaction constant which at
$J'_x \ll (J_x + J_y)$ is given by $ K = 1 + 2g + O(g^2), $ where $g
= J' _x (2) a_0/\pi v \ll 1$.
\medskip

Now we turn to $H' _2$. We have: \bea && a^{\dagger}_{n}
a^{\dagger}_{n+2} + h.c. \\ && \to - \left[ R^{\dagger}(x)
L^{\dagger} (x+\alpha) + L^{\dagger}(x) R^{\dagger} (x+\alpha) +
h.c. \right]\nn\\ && \,\, + (-1)^n \left[ R^{\dagger}(x) R^{\dagger}
(x+\alpha) + L^{\dagger}(x) L^{\dagger} (x+\alpha) + h.c. \right].
\nn \label{second} \eea Bosonizing the smooth term in the r.h.s. of
(\ref{second}) one obtains $\p_x \Phi \cos \sqrt{4\pi} \Theta$.
Bosonizing the staggered term yields $\sin \sqrt{4\pi} \Phi \cos
\sqrt{4\pi} \Theta $. Using the OPE
\begin{eqnarray*}
& &\sin \sqrt{4\pi} \Phi(x) \sin \sqrt{4\pi} \Phi(x+\alpha) \\
& & \quad \quad = {\rm const} - \pi \alpha^2 (\p_x \Phi)^2 -
\frac{1}{2} \cos \sqrt{16 \pi} \Phi,
\end{eqnarray*}
we find that,
in the continuum limit,
the Hamiltonian density ${\cal H}'_2$ is contributed by the operators
$\cos \sqrt{4\pi} \Theta$ and
$(\p_x \Phi)^2 \cos \sqrt{4\pi} \Theta$
(as before, we drop corrections related to  Umklapp processes).
The former leads to a small additive renormalization of the fermionic mass $\gamma$
and thus produces a shift of the critical point. The latter
represents an irrelevant perturbation (with scaling dimension 3) at the XX criticality.
In a noncritical regime it
renormalizes the mass and
four-fermion coupling constant $g$.
\medskip

Considering the structure of the remaining terms in the expansion (\ref{extra})
one arrives at similar conclusions. Here a remark is in order.
The only dangerous perturbation which would dramatically affect the above
picture is $\sin \sqrt{4\pi} \Theta$.
The presence of two nonlinear terms in the Hamiltonian,
$\gamma \cos \sqrt{4\pi} \Theta + \delta \sin \sqrt{4\pi} \Theta$,
would make the fermionic mass equal to $\sqrt{(\lambda - \lambda_c)^2 + \delta^2}$.
The Gaussian criticality in this case would never be reached, the model would always remain
massive, and nonzero staggered pseudospin densities, $\eta^z$ and $\eta^x$,
would coexist in the whole parameter range of the model.
\medskip

Fortunately, the appearance of the operator $\sin \sqrt{4\pi} \Theta$ is forbidden by
symmetry. The initial Hamiltonian (\ref{ham}) is invariant under
global pseudospin inversion In the $z$-component only: $\tau^z _n \to - \tau^z _n$.
After rotation $\tau^z \to \tau^y$
this translates to $\tau^y _n \to - \tau^y _n$.
Using the bosonized
expressions (\ref{stag_pseudospin}) for the staggered
pseudospin densities we find that the corresponding transformation of the dual field
is $\Theta \to \sqrt{\pi}- \Theta $ and so the bosonized Hamiltonian density
must be invariant under this transfomation.
This explains why the operator
$\sin \sqrt{4\pi} \Theta$ cannot appear in the
effective continuum theory.




\begin{thebibliography}{99}
\bibitem{footnote}
In the most general setting, there will be a contribution to the NMR
relaxation rate coming directly from the orbital sector. However, 
since the coupling to orbitals is non-local, this contribution is thus
less singular and is usually negligible
compared with contribution from spins.
\end{thebibliography}
\end{document}